\documentclass[12pt,preprint]{aastex}

\shorttitle{Dark Matter Structures}
\shortauthors{Kubo et al.}

\begin{document}


\title{Dark Matter Structures In The Deep Lens Survey}


\author{Jeffrey M. Kubo\altaffilmark{1}, Hossein Khiabanian\altaffilmark{2}, Ian P. Dell'Antonio\altaffilmark{2}, David Wittman\altaffilmark{3}, J. Anthony Tyson\altaffilmark{3}}


\altaffiltext{1}{Center for Particle Astrophysics, Fermi National Accelerator Laboratory, Batavia, IL 60510; kubo@fnal.gov}
\altaffiltext{2}{Physics Department, Brown University, Box 1843, Providence, RI 02912; ian@het.brown.edu, hossein@het.brown.edu}
\altaffiltext{3}{Physics Department, University of California, 1 Shields Avenue, Davis, CA 95616; dwittman@physics.ucdavis.edu, tyson@physics.ucdavis.edu}

\begin{abstract}
We present a regularized maximum likelihood weak lensing reconstruction of the Deep Lens
Survey F2 field (4 $\rm{deg^{2}}$).  High signal-to-noise ratio peaks in our lensing significance map appear to be
associated with possible projected filamentary structures.  The largest apparent
structure extends for over a degree in the field and has contributions from known optical clusters at three redshifts $(z\sim0.3,0.43,0.5)$.  Noise in weak lensing reconstructions is known to
potentially cause ``false positives''; we use Monte Carlo techniques to estimate the contamination in our sample, and find that 10-25\% of the peaks are expected to be false detections.  For significant lensing peaks we estimate the total signal-to-noise ratio of detection using a method that accounts for pixel-to-pixel correlations in our reconstruction.  We also report the detection of a candidate relative underdensity in the F2 field with a total signal-to-noise ratio of $\sim5.5$.  
\end{abstract}






\section{Introduction}
The mass function of galaxy clusters is a well known probe of cosmological
parameters \citep{haiman01}.  Since cluster mass is not a direct observable
other proxies such as optical galaxy richness \citep{koester07}, X-ray
temperature \citep{rosati02}, the Sunyaev-Zel'dovich decrement
\citep{carlstrom02}, or weak gravitational lensing \citep{schneider96} are
used to trace the mass distribution. To produce accurate constraints on
cosmological parameters the mass-observable conversion must be properly
calibrated in order to avoid potential biases.  With weak lensing, mass
overdensities are selected from maps of the projected mass density and peaks
of mass are identified as a function of their signal-to-noise ratio.  The observable
in this case is the weak lensing `shear' which is independent of baryonic
physics, and theoretically well understood.  This is a major strength that
weak lensing has over other methods.  Known limitations of this method are
projection effects due to large scale structure \citep{hennawi05,white02}
which can produce spurious detections, a low peak detection significance,
complicated point-spread function (PSF) corrections, and the observational
expense (since it requires deep imaging).  In spite of these limitations,
forecasts for future imaging surveys predict that this method will still
provide large samples of shear-selected galaxy clusters which will allow for
constraints on cosmological parameters, including dark energy
\citep{marian06}. In addition to individual clusters, studies have also
pointed out that the global statistics of dark matter peaks in weak lensing
maps can be used to put constraints on cosmology \citep{jain00}.  Within a
given dataset cosmological information from peak statistics could also play a
complementary role in breaking degeneracies in cosmological parameters from
studies of cosmic shear \citep{gavazzi07}. 

The current generation of deep optical imaging surveys are creating the first
ever maps of the dark matter distribution with weak lensing.  In
\citet{wittman06} we presented preliminary maps from the Deep Lens Survey
(DLS), covering a total of area of $8$ $\rm{deg}^{2}$ spread over five fields.
Initial maps from the Canada-France-Hawaii Telescope Legacy survey covering
four separate $1$ $\rm{deg}^2$ fields were presented in \citet{gavazzi07}.
Recently, results from the Hubble Space Telescope COSMOS 2 $\rm{deg}^2$ field
were reported in \citet{massey07}.  To reconstruct the projected mass density
($\kappa$) in these surveys from weak lensing, all of these studies have
implemented a `direct' reconstruction method based on or related to the technique of
\citet{kaiser93}.  This reconstruction method is computationally efficient but
has a number of known limitations; for instance, it requires setting an
arbitrary smoothing scale, and it is difficult to incorporate other lensing
effects such as magnification \citep{seitz98}.  Several different
reconstruction techniques have been developed to overcome these issues, the
most promising of which advocate reconstructing the deflection potential
($\psi$) on a grid rather than directly determining $\kappa$.  These
techniques have primarliy been applied to pointed observations of
individual clusters, for instance in \citet{jee07} or \citet{bradac06}. 
In this study we use a regularized maximum likelihood technique to reconstruct the projected
mass distribution over a wide area, four square degree field from the DLS. 

Our paper is organized as follows: In $\S \ref{sec:data}$ details of the imaging data, PSF correction, and the source galaxies used in our analysis are presented. In $\S \ref{sec:analysis}$ the maximum likelihood
algorithm used in our reconstruction and its application to the DLS are discussed. In $\S
\ref{sec:masspeaks}$ the structures observed in the lensing reconstruction are presented along with their likely optical counterparts.  In $\S \ref{sec:discussion}$ the expected false peak rate due to lensing
shape noise is discussed as well as our method of determining the total signal-to-noise ratio
of lensing peaks.  In $\S \ref{sec:summary}$ our results are summarized and we discuss future directions of our work. 

\section{Data}
\label{sec:data}

The DLS is a deep imaging survey of five widely separated four square
degree fields (F1-F5) in $BVRz'$.    Fields in the DLS were selected in a
blind manner the only restrictions were to avoid bright foreground galaxies,
known low redshift clusters ($z\lesssim0.1$), and areas of high extinction
\citep{wittman02}.  A primary motivation for this was to create an unbiased
shear selected sample of clusters.  In this study we restrict our analysis to
the F2 field centered on $\rm{R.A.}=09^{\rm{h}}19^{\rm{m}}32.4^{\rm{s}}$,
$\rm{decl.}=+30^{\circ}00\arcmin00\arcsec$.  Observations of the F2 field were
carried out using the MOSAIC I imager \citep{muller98} on the Kitt Peak Mayall
4m telescope.  Observations of F2 began at Kitt Peak in 1999 November and
ended in 2004 November.  This was the first DLS field and to have complete
imaging.  The observing strategy for the DLS is to split each field into a
$3\times3$ grid of $40\arcmin\times40\arcmin$ subfields.  Each completed
subfield consists of twenty 900s $R$ band exposures, and twenty 600s exposures
in each of the $BVz'$ filters.  The $R$ band is used as the primary filter to
measure galaxy shapes for weak lensing, and observations were only carried out
in this filter when the seeing FWHM was $<0.9\arcsec$.  The remaining filters
($BVz'$) are primarily used to measure photometric redshifts of source
galaxies, however these are not used in our current analysis. 

For our lensing analysis the shapes of background galaxies are measured from
the co-added $R$ band subfield frames.  A detailed discussion of the DLS weak
lensing pipeline is given in \citet{wittman06} here we provide a brief
overview. Basic reductions such as flat fielding and bias are performed using
the IRAF package MSCRED.  The MOSAIC cameras each consist of eight individual
CCD detectors, and the PSF in each detector is determined by identifying stars
from the stellar locus in magnitude-size space.  We fit the spatial variation
of the PSF in each detector using a third order polynomial, and a $3\times3$
rounding kernel is used to circularize the PSF as in \citet{fischer97}.  The
PSF corrected $R$ band images of each subfield are co-added using our custom
software DLSCOMBINE \citep{wittman06}.  The final co-added subfield images
reach a depth of $R\sim26$. 

To initially detect objects in each subfield image we use the SExtractor
 \citep{bertin96} package.  Shapes for detected objects are then remeasured
 with adaptive moments using ELLIPTO \citep{smith01,heymans06}, which is a
 partial implementation of the algorithm of \citet{bernstein02}.  Objects
 which triggered error flags in ELLIPTO were removed since these are objects
 where adaptive moments did not converge or indicated a problem with shape
 measurement.  We also rejected objects which triggered SExtractor flags
 indicating a fatal error occured during shape measurement, the object contained a saturated pixel, or the object was too close the edge of the image.  For our analysis we use source galaxies in the magnitude range $22.0<R<25.5$, where the number counts in the field peak at $R\sim25.5$.  We include only galaxies with ELLIPTO size $>1.25\times \rm{PSF}$ at the
 position of each galaxy.  The maximum ELLIPTO size is set at $20.0$
 $\rm{pixel^{2}}$ in order to eliminate potential low redshift, low surface brightness galaxies from
 our sample.  Our size and magnitude cuts produce a sample of source galaxies that each have a signal-to-noise ratio $\gtrsim 15$.  We also reject objects whose observed ellipticity is $e>0.6$, as
 these are likely superpositions of objects \citep{wittman00}.  The dilution (or seeing) correction is performed using simulations described in $\S \ref{sec:reg}$.  This is chosen over the analytic dilution correction described for the ELLITPO method in \citet{heymans06}, since it has been pointed out that this correction does not correctly incorporate the kurtosis of the PSF \citep{hirata03}. The final
 subfield catalogs were stitched together to create a supercatalog of source
 galaxies for the F2 field.  This catalog contains $\sim 328,000$ galaxies
 over the field, or $\sim 23$ galaxies $\rm{arcmin^{-2}}$.  

 Our current source galaxy catalog is not the same catalog used in \citet{wittman06} or \citet{hossein08}.  In both of these studies the catalog was restricted to brighter galaxies $(R<25)$ with larger galaxy sizes than we used here.  This work uses a catalog with a factor of two more source galaxies than was used in \citet{hossein08}.





\section{Weak Lensing Analysis}
\label{sec:analysis}
\subsection{Formalism}
In the thin lens approximation the mapping of light from the source plane
$\vec{\beta}$ to image plane $\vec{\theta}$ is given by the lens equation
$\vec{\beta}=\vec{\theta}-\vec{\alpha}(\vec{\theta})$.  The deflection angle
$\vec{\alpha}$ of light due to a two dimensional Newtonian deflection
potential $\psi$ of a lens is given by 
\begin{equation}
\vec{\alpha}(\vec{\theta})=\vec{\bigtriangledown}\psi.
\end{equation}
The surface mass density ($\kappa$) of the lens can be calculated directly from the deflection potential by
\begin{equation}
\label{eqn:kappa}
\kappa=\frac{\Sigma}{\Sigma_{crit}}=\frac{1}{2}(\psi_{11}+\psi_{22}),
\end{equation}
where subscripts on $\psi$ refer to the partial derivatives $\psi_{ij}\equiv\frac{\partial^{2}\psi}{\partial\theta_{i}\partial\theta_{j}}$.  Here the critical surface mass density ($\Sigma_{crit}$) is given by
\begin{equation}\Sigma_{crit}=\frac{c^{2}}{4\pi G}\frac{D_{s}}{D_{d}D_{ds}},\end{equation}
where $D_{d}$, $D_{s}$, and $D_{ds}$ are the angular diameter distances to the
lens, source, and between the lens and source respectively.  The tidal
gravitational field of the lens, or shear, is described in complex notion by
$\gamma=\gamma_{1}+\rm{i}\gamma_{2}$ where the shear components are given by 
\begin{equation}
\begin{array}{c}
\gamma_{1}=\frac{1}{2}(\psi_{11}-\psi_{22})\\\\
\gamma_{2}=\psi_{12}=\psi_{21}\\
\end{array}
\end{equation}
and the complex reduced shear is given by $g=\gamma/(1-\kappa)$
\citep{schneider00}.  For a more detailed discussion of lensing we refer the
reader to \citet{kochanek05}. 

\subsection{Regularized Maximum-Likelihood Reconstruction}
\label{sec:reg}
Maximum likelihood methods for weak lensing cluster mass reconstruction were
first demonstrated in \cite{bartelmann96}.  Several variants of this algorithm
have since been proposed, for instance the methods described in
\cite{bridle98} and \cite{seitz98}. To produce a weak lensing convergence map
of the F2 field we use a regularized maximum likelihood approach based on the method
developed by \citet{seitz98} with improvements described in \citet{hossein08}.
In this technique the deflection potential in the field is determined over a
grid, and the convergence map is generated directly from the deflection
potential map using equation $(\ref{eqn:kappa})$. 

To produce a convergence map with dimensions $N_{x}\times N_{y}$, a grid of
the deflection potential with dimensions $(2N_{x}+4)\times(2N_{y}+4)$ is used.
The twice larger grid is required here to fix the ringing effects in the
projected mass maps caused by second order numerical differentiation of the
deflection potential.  The extra rows and columns are needed to compute
$\kappa$ and $\gamma$ using second order finite differencing.  With this
method we minimize the function $\mathcal{F}$, 
\begin{equation}
\mathcal{F}=\frac{1}{2}\chi^{2}+\lambda\mathcal{R},
\label{eqn:F}\end{equation}
where $\lambda$ is the regularization coefficient, and $\mathcal{R}$ is the
regularization function.  The $\chi^{2}$ term is determined from
\begin{equation}
\chi^{2}=\frac{1}{N_{g}}\sum_{k=1}^{N_{g}}\frac{(\epsilon_{k}-<\epsilon>(\textbf{x}_{k}))^2}{\sigma_{\epsilon}},
\end{equation}
where $N_{g}$ is the number of source galaxies, $\epsilon_{k}$ is the complex
ellipticity of a galaxy at position $x_{k}$, and  $<\epsilon>$ is the expected
ellipticity at $x_{k}$.  We compute the expected ellipticity distribution and
the dispersion $\sigma_\epsilon$, as a function of reduced shear from
simulations \citep{hossein08}.  The ellipticity
distribution in our simulations is based on the measured distribution from the
Hubble Ultra Deep Field \citep{beckwith06}. We regularize $\chi^{2}$ using a
zeroth-order regularization function given by 
\begin{equation}
\mathcal{R}=\sum_{m=1}^{N_{x}}\sum_{n=1}^{N_{y}}(\kappa_{mn}-p_{mn})^{2}
\label{eqn:R}\end{equation}
where $\kappa_{mn}$ is the convergence at a grid point and $p_{mn}$ is the
prior.  This form of the regularization function is chosen for simplicity, but
also ensures the smoothness of the reconstruction.  The regularization
coefficient, $\lambda$, in $\mathcal{F}$ represents a compromise between the
best fit $\chi^{2}$ and the closest match to the prior.  We
minimize the function $\mathcal{F}$ using the conjugate gradient method from
\citet{press92}. 

\begin{figure}
\epsscale{1.0}
\plotone{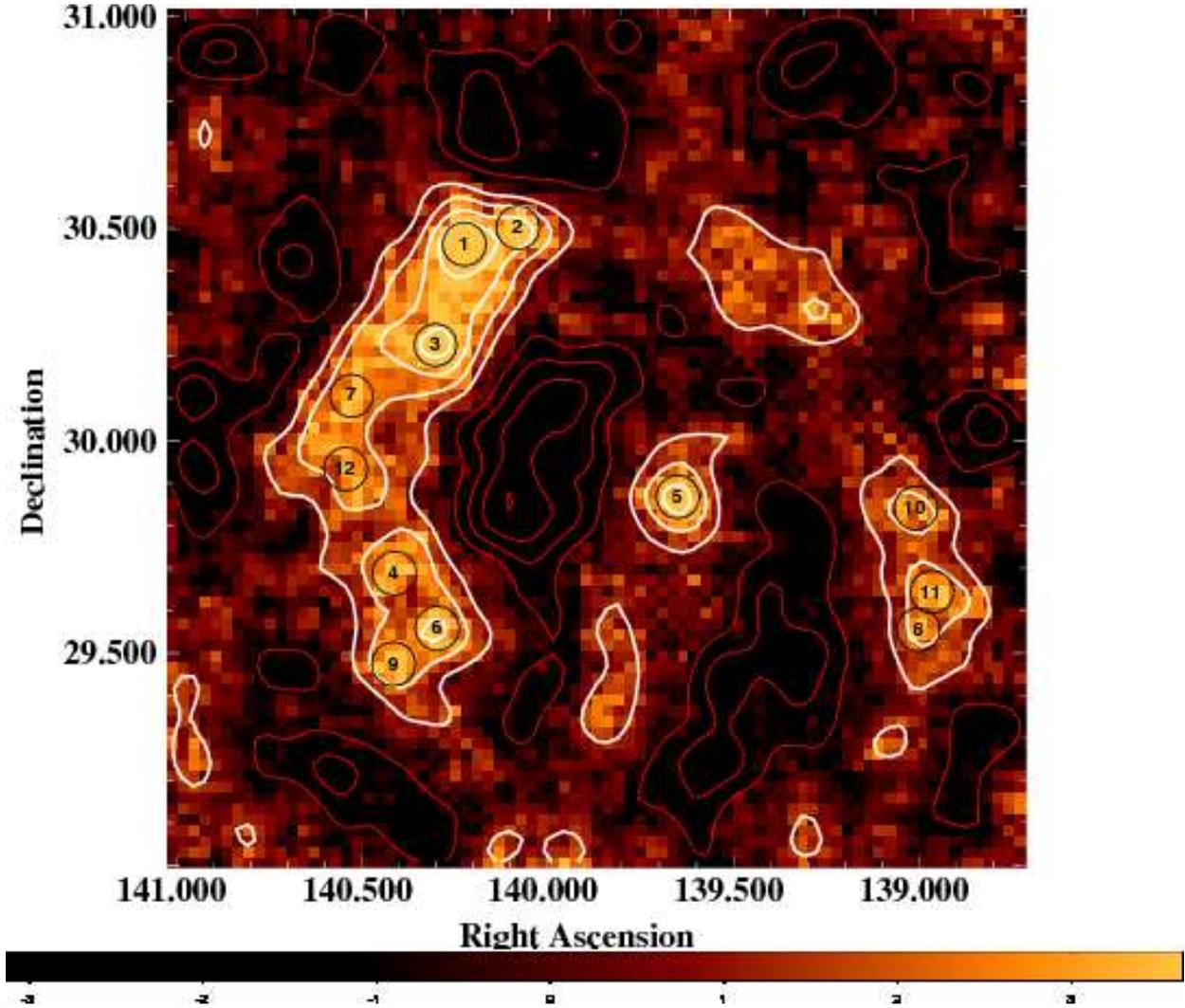}
\caption{Weak lensing $\kappa-\rm{S/N}$ map of the F2 field centered on $\rm{R.A.}=09^{\rm{h}}19^{\rm{m}}32.4^{\rm{s}}$, $\rm{decl.}=+30^{\circ}00\arcmin00\arcsec$.  Postive contours of signal-to-noise ratio (white) are overlaid and enclose the $4,2,3,$ and $1\sigma$ regions; negative contours (red) enclose the $-4,-3,-2,$ and $-1\sigma$ regions.  Black circles indicate the positions of the most significant peaks ($\nu>3.5$) with the rank of each peak shown at the center of each circle.}
\label{fig:snmap}
\end{figure}

Our reconstruction proceeds at a series of different resolutions, beginning
with a coarse $20\times20$ grid of the potential and a completely uniform
prior.  This outputs a coarse potential which is used to create a convergence
map.  The resulting convergence map is smoothed and used as a prior to the
next level of resolution.  Using a smoothed prior has been shown to produce
more accurate reconstructions \citep{seitz98,lucy94}.  As described in
\citet{hossein08} we choose the regularization coefficient at each of the
higher resolution reconstructions by minimizing $\mathcal{F}$
(eq. \ref{eqn:F}) with multiple values of $\lambda$ (between 0 and 10) along
with minimizing only $\mathcal{R}$ (eq. \ref{eqn:R}). We scale $\chi^2$ to
values between 0 and 1, using its lowest value obtained when $\lambda=0$ and
its highest value obtained when only $\mathcal{R}$ is minimized. Similarly, we
scale $\mathcal{R}$ to values between 0 and 1. The intersection of the scaled
$\chi^2$ vs $\mathcal{R}$ curve and the scaled line $\chi^{2}=\mathcal{R}$
determines the proper value of the regularization coefficient.  This processes
of producing maps at a given resolution and determining the proper
regularization coefficient is repeated for three resolutions of the convergence map
$(20\times20$, $40\times40$, $80\times80$), until the final map resolution is
achieved.  Our final convergence map has dimensions of $80\times80$ with a
plate scale of $1\arcmin.5$ $\rm{pixel^{-1}}$.  The resulting signal-to-noise ratio
map (convergence map divided by the rms map) is shown in Figure
\ref{fig:snmap}.  The rms map ($\kappa_{\rm{rms}}$) is described further
in $\S \ref{sec:peaks}$.

\section{Dark Matter Peaks in F2}
\label{sec:masspeaks}
\subsection{$\kappa$ Signal-To-Noise Ratio Map}
\label{sec:peaks}
We search for lensing peaks in the F2 field by first creating a $\kappa$
signal-to-noise ratio map (henceforth $\kappa-\rm{S/N}$ map).  As in
\citet{miyazaki07} we construct a $\kappa_{\rm{rms}}$ map using 100
Monte Carlo realizations of the original source galaxy catalog.  We note here that half as many realizations could have been used and the following results would not have changed.

\begin{figure}
\epsscale{1.0}
\plotone{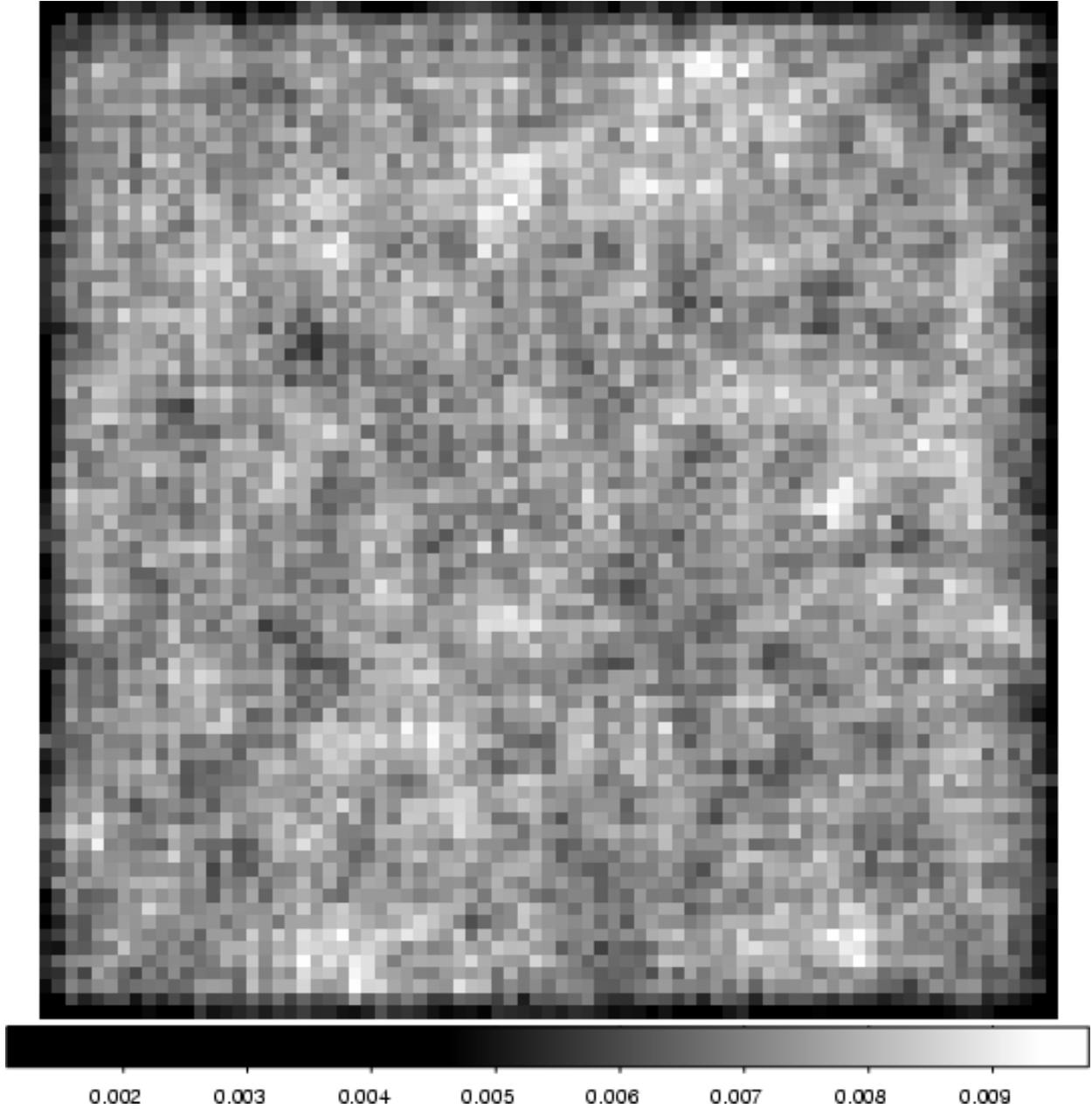}
\caption{The $\kappa_{\rm{rms}}$ noise map based on 100 Monte Carlo
  realizations of the F2 field.  Each individual Monte Carlo map is created by
  shuffling the source galaxy catalog positions and ellipticity components.  A
  maximum likelihood $\kappa$ map is made for each individual Monte Carlo
  realization.  Each pixel in the $\kappa_{\rm{rms}}$ map represents the
  $1\sigma$ error at this point over the set of Monte Carlo reconstructions.}
\label{fig:rms}
\end{figure}

 In each 
realization, the position and
ellipticity components of each galaxy are decoupled and randomly assigned to new galaxies.
Shuffling the source catalog in this manner is useful since it preserves the
real variation in background galaxy source density.  For each shuffled source
catalog, a maximum likelihood $\kappa$ map is created using the same
regularization coefficients at each resolution as was used to create our
original $\kappa$ map \citep{hossein08}.   From the resulting set of 100 Monte Carlo maps we
created the $\kappa_{\rm{rms}}$ map shown in Figure \ref{fig:rms}.  Each pixel
in this map represents the $1\sigma$ standard deviation at this point over the
set of Monte Carlo reconstructions.  A $\kappa-\rm{S/N}$ map for F2 is made by
dividing the original $\kappa$ map by the $\kappa_{\rm{rms}}$ map (Figure
\ref{fig:snmap}).  Contours in the $\kappa-\rm{S/N}$ map range from a peak
signal-to-noise ratio ($\nu$) of $1.0-6.0$; negative contours have been omitted here
for clarity. 

We detect peaks in the $\kappa-\rm{S/N}$ map using the SExtractor package
\citep{bertin96}.  Peaks are separated by setting the minimum contrast
parameter in SExtractor to 0 which separates all possible peaks.  We measure
peaks relative to a zero background level and use the height of each peak to approximate the signal-to-noise ratio of detection.  We set the minimum area for detection at 9 contiguous pixels and the minimum
threshold for detection at 0.01.  Our definition of a peak is similar to that given in Jain \& Van Waerbeke (200), however in our case our threshholding criteria eliminate many low-significance peaks based on size and significance.  This use of a minimum threshold provides a cleaner peak catalog because peaks below this level of significance do not correspond to real detections.  

The edge pixels in our $\kappa-\rm{S/N}$ map correspond to the edge of the F2 field, where the
imaging is shallower and the noise is higher.  Therefore, there are fewer
galaxies that correspond to these pixels.  Because the Monte Carlo procedure
preserves the spatial distribution of galaxies, these pixels are
systematically underconstrained.  As a result the variance along the edge of $\kappa_{\rm{rms}}$ map is smaller than in the rest of the field.  This can be seen as the dark border in Figure \ref{fig:rms}.  We find this causes spurious detections along the edge of the $\kappa-\rm{S/N}$ map, so to remove these we filter out a border region of size $3\arcmin$ from the lensing peak catalog.  

\begin{figure}
\epsscale{1.0}
\plotone{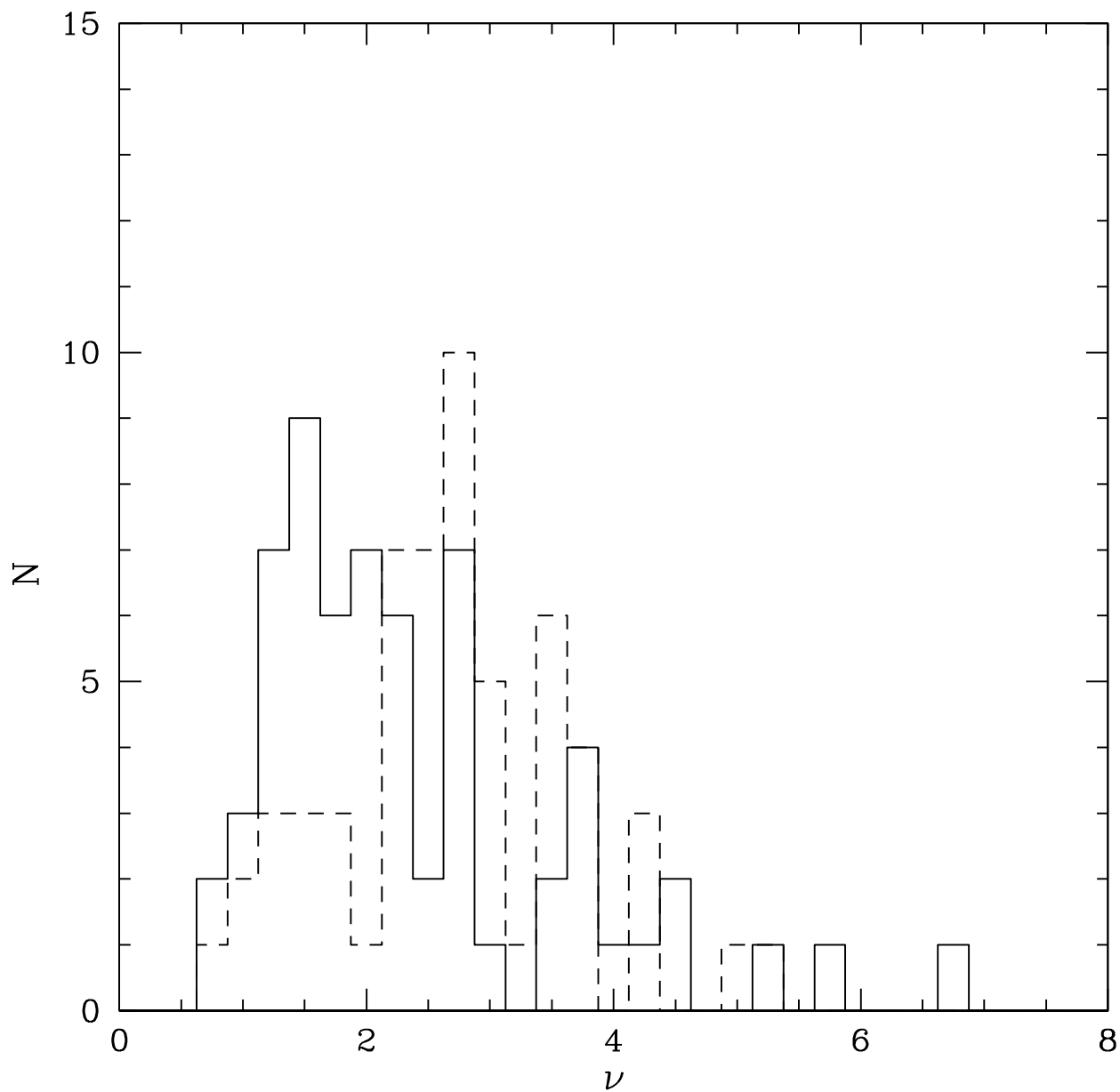}
\caption{Distribution of peak signal-to-noise ratio $(\nu)$ measured from our $\kappa-\rm{S/N}$ map of the DLS F2 field.  The solid histogram is the distribution of positive signal-to-noise ratio peaks, the dashed histogram is the distribution of negative signal-to-noise ratio peaks.  Negative values have been flipped here to the positive axis.}
\label{fig:peaks}
\end{figure}

The resulting peak distribution in F2 as a function of $\nu$ is shown in Figure
\ref{fig:peaks}.  The solid black histogram shows the distribution of positive
peaks, the most significant being the confirmed cluster Abell 781.  Unlike
previous studies \citep{gavazzi07} our use of an effective peak threshold cuts
off the low signal-to-noise ratio end of the peak distribution, however below this level peaks do not correspond to real detections.  The dashed histogram is the distribution of
negative peaks where the $\kappa-\rm{S/N}$ map has been flipped to the
positive axis for comparison.  The physical interpretation of significant
negative peaks are underdensities in the matter distribution
\citep{miyazaki02}.  A possible underdensity is detected in the DLS F2 field,
which we discuss further in $\S \ref{sec:void}$. 

\subsection{High S/N Peaks}
\label{sec:clusters}

To be consistent with previous studies we restrict our study of lensing peaks
in the F2 field to peaks with $\nu>3.5$ \citep{gavazzi07}.  Above this signal-to-noise ratio we detect 12 peaks, with positions and
$\nu$ values listed in Table \ref{tab:clusters}.  All of these high
signal-to-noise ratio peaks appear to lie along projected structures resembling
filaments in the $\kappa-\rm{S/N}$ map shown in Figure \ref{fig:snmap}.
Positions of peaks are overlaid in circles on the $\kappa-\rm{S/N}$ map, with
the rank of each peak (sorted by peak value) shown in each circle.  Three
distinct projected structures appear in the reconstruction---the largest
projected structure spanning over a degree.  Below we discuss each of these
structures and their association with likely optical counterparts in the
imaging data. 

\subsubsection{Eastern Structure}
In the Eastern section of our map we detect a projected structure consisting
of eight significant peaks which extend along the North-South direction and
span for over a degree.  The two highest significance peaks (Peaks 1 \& 2) are
associated with the cluster complex Abell 781, of which we reported a previous
lensing detection in \citet{wittman06}.  X-ray observations of this system
\citep{sehgal08} have shown this complex consists of four separate clusters.
In our lensing reconstruction we have separated two distinct clumps, where the
easternmost clump in our map (Peak 1) is the most significant ($\nu=6.6$).
Due to the pixel scale in our reconstruction ($1\arcmin.5$ $\rm{pixel}^{-1}$)
this clump is detected as a superposition of two clusters, CXOU J092110+302751
at $z=0.427$, and CXOU J092053+302800 at $z=0.291$.  Our previous direct
lensing reconstruction in \citet{wittman06} referred to these as clumps C and
B respectively.  To the West of this peak we detect our second highest ranked
peak (Peak 2 with $\nu=5.7$) which is the cluster CXOU J092026+302938 at
$z=0.302$.  This peak was referred to as clump A in \citet{wittman06}.
\citet{sehgal08} recently reported the X-ray detection of an additional
cluster to the west of the A781 complex (XMMU J091035+303155 at $z=0.428$)
which has a lensing significance of $1\sigma-2\sigma$ when fit to a
Navarro, Frenk, \& White (NFW) profile \citep{navarro96}.  In our maximum
likelihood reconstruction we note that we are not able to detect this
cluster in our map.  As we discuss in $\S \ref{sec:a781}$, more work is
needed to explain the low detection significance of this cluster.  Images of
the Abell 781 complex can be found in \citet{wittman06}. 

South of the Abell 781 complex along the Eastern structure we detect a new
significant peak (Peak 3) with $\nu=5.3$.  The DLS optical imaging reveals a
large excess of galaxies located near the peak center.  Photometric redshifts
of likely cluster members taken from the SDSS Data Release 6 (DR6) photoz2
table \citep{oyaizu08} place this cluster at a redshift of $z\sim0.5$.  We
stress that this is only a preliminary redshift based on a small number
$(\sim5)$ of likely cluster members.  This lensing peak likely has
contributions from other clusters along the line of sight; for instance, we
note that the SDSS Maxbcg cluster catalog \citep{koester07} which overlaps the
F2 field also detects another cluster with optical galaxy richness
$N_{\rm{gals}}=15$ and redshift $z=0.297$ located $\sim 3\arcmin.8 $ to the
southeast of this peak position.  At this angular separation however, this cluster does not likely
contribute much signal to this lensing peak.

South of this peak we detect two other significant peaks in the Eastern
structure, Peaks 7 and 12, with $\nu=4.1$ and $\nu=3.6$ respectively.  The DLS
optical imaging reveals several galaxy overdensities in the vicinity of peak which likely contribute to the lensing signal in each.  We currently do not have richness or redshift information for the groups, so cannot disentangle the relative contributions to each lensing peak.  South of this grouping we detect another grouping of three peaks : Peaks 4 ($\nu=4.5$), Peak
6 ($\nu=4.2$), and Peak 9 ($\nu=3.9$), where the fourth ranked peak is north
of the two other lensing peaks.  From the DLS imaging Peaks 4 and 6 do not appear to correspond to galaxy overdensities and are potentially false positives. Peak 9 is likely associated with many optical overdensities; for instance, $\sim1.5\arcmin$ to the southwest of peak 9 there is a group scale system which has a single spectroscopic measurement from the SDSS DR6 of $z=0.339\pm0.0002$ \citep{adelman08}.
 
The Eastern Structure appears to be a superposition of structures lying at several different redshifts ($z\sim0.3,
0.43, 0.5$).  More follow-up work is needed however to disentangle the relative
strengths of the lensing contributions at each redshift.  The Smithsonian
Hectospec Lensing Survey (SHELS) \citep{geller05}, a redshift survey being
conducted in the F2 field, should provide further insight into this
structure. 

\subsubsection{South-West Structure}

In the southwest portion of our map we detect another structure containing
three significant peaks.  Two of these peaks (Peaks 8 \& 11) are associated
with the system DLSCL J0920.1+3029 which we previously reported in
\citet{wittman06}.  This system consists of two known clusters separated by
$\sim 9\arcmin$, which each have a spectroscopic redshift of $z=0.53$. 
The northern peak (11th ranked peak) is detected with $\nu=3.7$ and
the southern peak (8th ranked peak) is detected with $\nu=3.9$.
Images of the clusters associated with peaks 8 and 11 can be
found in \citet{wittman06}. 

North of these peaks we detect a new peak (Peak 10) with $\nu=3.8$.  This
lensing peak lies near an overdensity of galaxies $\sim 3\arcmin$ to the East centered on $\rm{R.A.}=09^{\rm{h}}16^{\rm{m}}15.7^{\rm{s}}$, $\rm{decl}=+29^{\circ}50\arcmin05.5\arcsec$.  A $\sim 3\arcmin$ separation between the center of the lensing peak and the optical overdensity is large and could possibly indicate that this peak is spurious, however the broad lensing contours do encompass this galaxy overdensity.  Photometric redshifts of likely cluster members from the SDSS DR6 photoz2 table \citep{oyaizu08} place this cluster at a redshift of $z\sim0.55$.  This
redshift is also only an estimate based on a small number $(\sim5)$ of likely
cluster member galaxies.  This cluster and the other two lensing peaks likely
place the redshift of much the southwest structure at $z\sim0.5$, however
other structures along the line of sight could also be contributing to the
lensing signal here. 

\subsubsection{Central Peak}

\begin{figure}
\epsscale{1.0}
\plotone{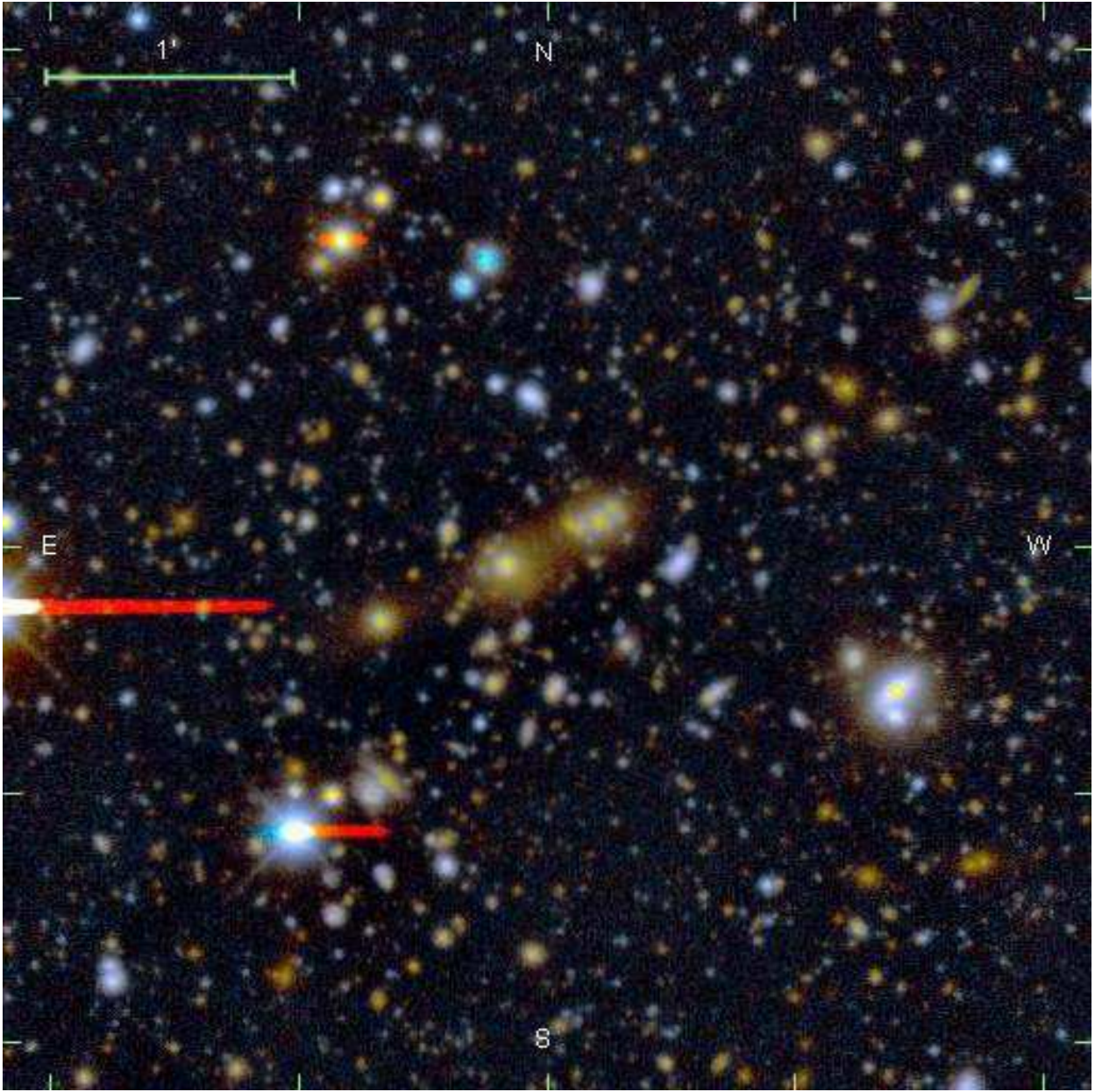}
\caption{Galaxy cluster coincident with Peak 5 in our high signal-to-noise ratio peak sample.  North is up, East is to the left.  The cluster candidate contains two bright elliptical galaxies, the eastern-most having a spectroscopic redshift of $z=0.3171$ from the SDSS DR6.}
\label{fig:peak5}
\end{figure}

Near the center of our map we detect a new significant lensing peak (Peak 5) with
$\nu=4.4$.  The optical imaging reveals two brightest cluster galaxies (BCG's)
coincident with this peak, surrounded by an excess of fainter red
galaxies (Figure \ref{fig:peak5}).  The BCG located at
$\rm{R.A.}=09^{\rm{h}}18^{\rm{m}}36^{\rm{s}}.1$,
$\rm{decl}=+29^{\circ}53\arcmin08\arcsec$ has a spectroscopic redshift from
the SDSS DR6 of $z=0.3171\pm0.0002$ \citep{adelman08}.  We note that the SDSS
Maxbcg cluster catalog detects a cluster $\sim 2\arcmin.5$ to the East of this
peak with richness $N_{\rm{gals}}=17$ at a slightly lower redshift $z=0.278$,
which could also contribute to the lensing signal. 

\subsection{Low S/N Peaks}

We comment here on lower S/N peaks which appear to be associated with the
projected structure seen in the lensing reconstruction, but do not fall in our
high S/N sample.  South of our 5th ranked peak, the $\kappa-\rm{S/N}$ map
(Figure \ref{fig:snmap}) reveals a structure which we detect as two peaks; the
first located at $\rm{R.A.}=09^{\rm{h}}19^{\rm{m}}26.1^{\rm{s}}$,
$\rm{decl.}=+29^{\circ}22\arcmin06.4\arcsec$ with $\nu=2.7$ and the second
located at $\rm{R.A.}=09^{\rm{h}}19^{\rm{m}}19.0^{\rm{s}}$
$\rm{decl.}=+29^{\circ}33\arcmin04.2\arcsec$ with $\nu=2.4$. 

North of Peak 5 we detect another structure in the $\kappa-\rm{S/N}$ map which
are resolved as three separate peaks, which we designate as A, B, \& C.  Peak
A is at $\rm{R.A.}=09^{\rm{h}}20^{\rm{m}}56.5^{\rm{s}}$,
$\rm{decl.}=+30^{\circ}16\arcmin36.6\arcsec$ with $\nu=2.7$, Peak B at
$\rm{R.A.}=09^{\rm{h}}20^{\rm{m}}55.0^{\rm{s}}$,
$\rm{decl.}=+30^{\circ}08\arcmin44.0\arcsec$ with $\nu=2.7$, and Peak C at
$\rm{R.A.}=09^{\rm{h}}20^{\rm{m}}02.9^{\rm{s}}$,
$\rm{decl.}=+30^{\circ}06\arcmin34.6\arcsec$ with $\nu=3.4$.  The superpositions of these clusters as
well as potentially other clusters along the line of sight likely contribute
to the lensing peaks in this structure. 

\section{Discussion}
\label{sec:discussion}

\subsection{False Peak Rate}
The noise in weak gravitational lensing reconstructions is influenced by
several sources: the distribution in background galaxy positions,
ellipticities, and redshifts, spatial variations in the sky background and PSF,
and pixel-to-pixel correlations due to the reconstruction methods.  As a
result, the noise distribution in the lensing maps is strictly non-Gaussian,
and there is the potential for ``false positives''---peaks in the lensing map
that do not correspond to any real objects. 

Several previous studies have addressed the issue of the false peak rate in weak lensing mass reconstructions.  \citet{hamana04} used N-body simulations and ray-tracing to construct catalogs of expected clusters to compare with them.  They reported that the completeness and contamination of the catalogs depended on the S/N threshold, and that even at relatively high thresholds ($\sim 5\sigma$), a 20-30\% incompleteness and a similar contamination rate would be expected. \citet{schirmer07} used a variation of the aperture mass statistic \citep{schneider96} and estimates of the projected galaxy overdensity to estimate the noise contamination rate in a sample of ESO-WFI fields.  When they varied the S/N threshold, they found contamination rates of 30-80\%, depending on the depth and seeing of the fields.  \citet{hetterscheidt05} applied two variations of the aperture mass statistic to 50 deep Very Large Telescope fields; they set their thresholds so as to approximate a 20\% contamination rate from noise peaks in their sample.  Even though these studies used very different reconstruction methods, data sets and selection criteria, they all concur in finding 20-40\% contamination in the rough S/N range that we consider.  At lower S/N, the very large number of ``detections" in both $\kappa$ and randomized maps indicate a much higher contamination rate.  We also mention that \citet{vanwaerbeke00} studied the noise in weak lensing reconstructions, however their analysis is not directly applicable to us since it does not apply to maximum likelihood reconstructions which use a regularization term.

\begin{figure}
\epsscale{1.0}
\plottwo{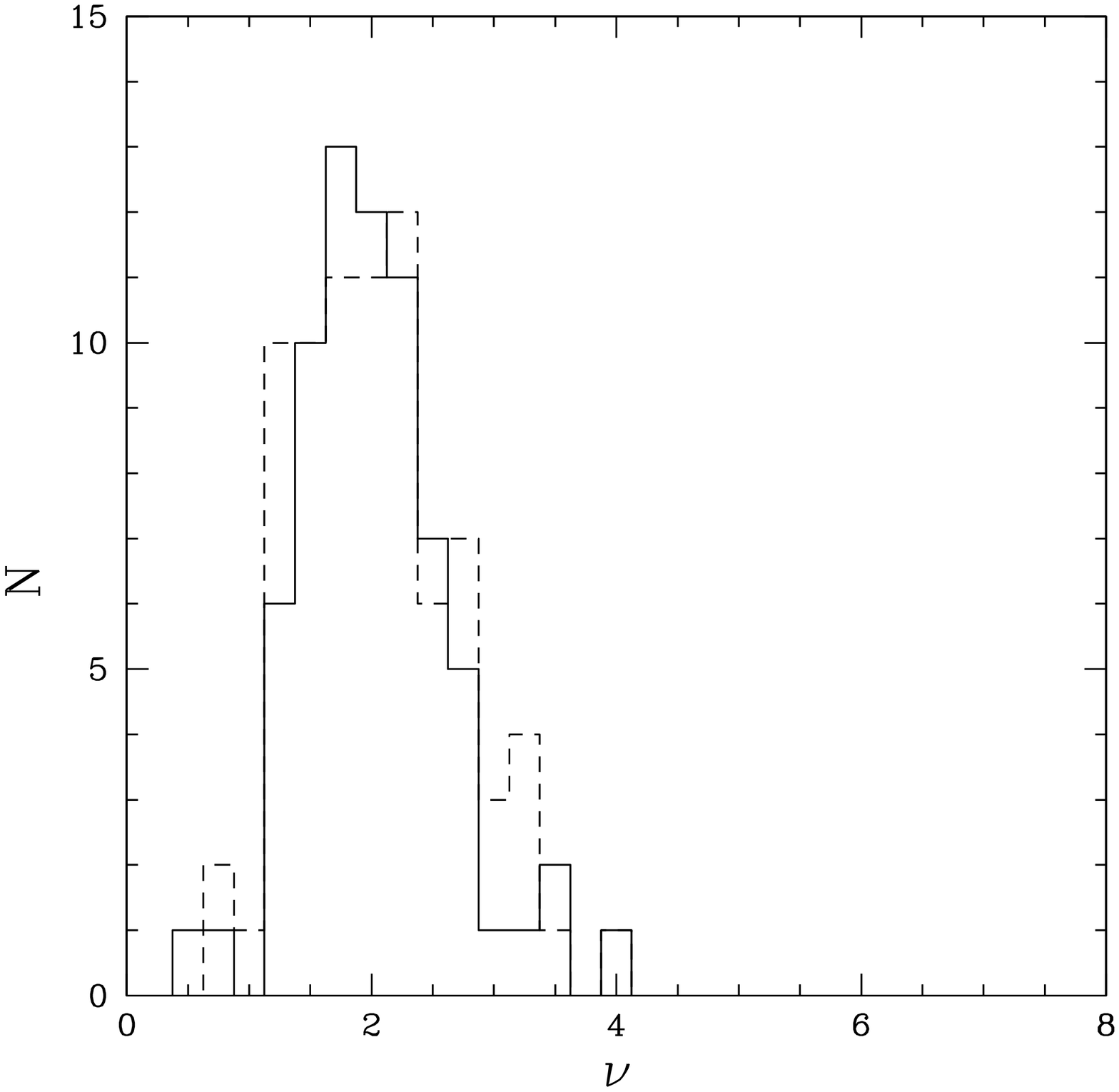}{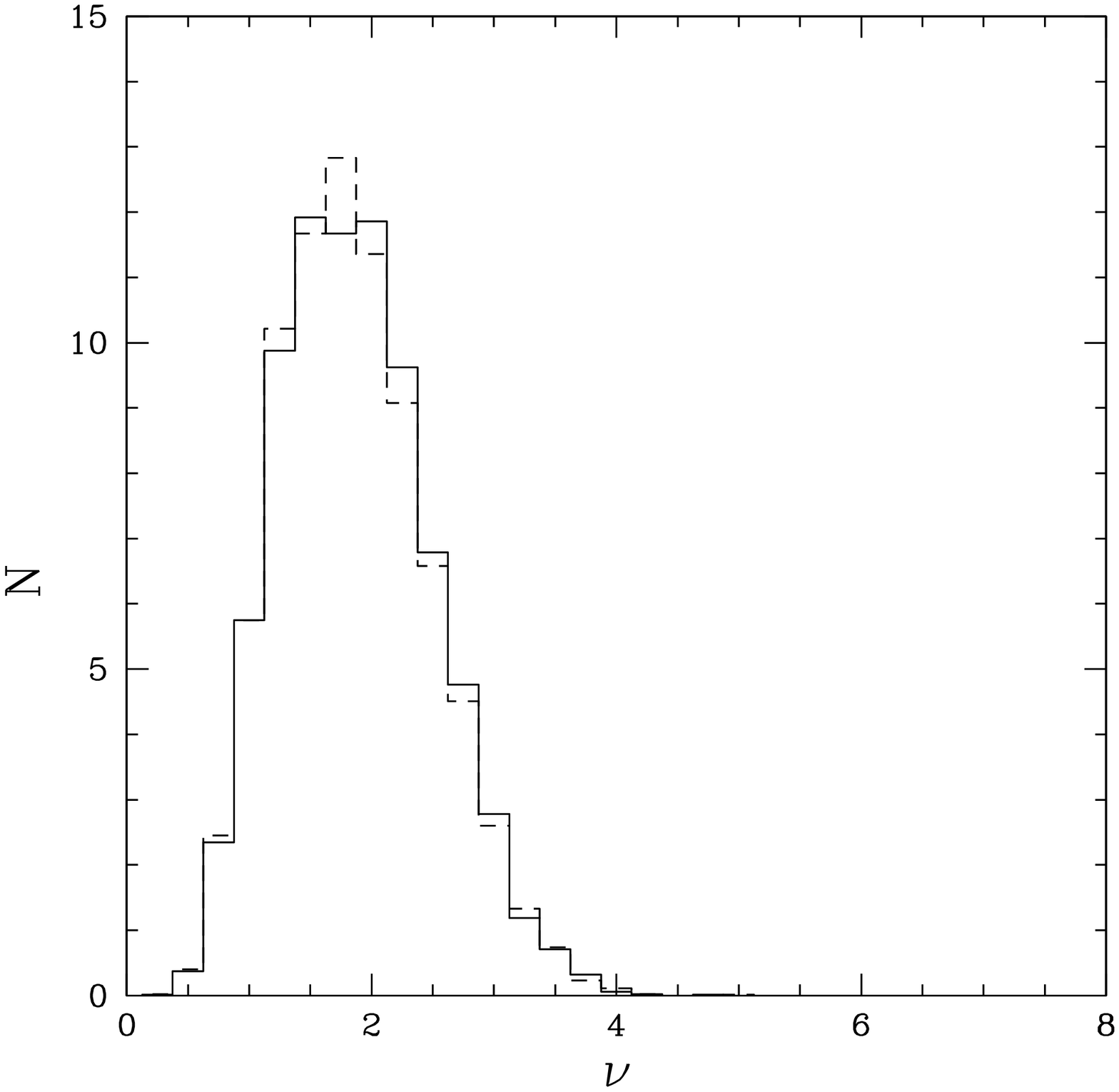}
\caption{Distribution of peak signal-to-noise ratio ($\nu$) for the mean Monte Carlo map (Left) and for the average of the individual Monte Carlo maps (Right).  As in our real lensing map solid histograms are positive peaks, dashed histograms are negative peaks (where negative values have been flipped to the positive axis).  For our high signal-to-noise ratio cut $(\nu>3.5)$ three peaks are recovered in the mean map, $\sim 1$ is recovered in the average of the individual Monte Carlo maps.  These tests provide an estimate of the number expected false peaks in our high signal-to-noise ratio sample due to lensing shape noise.}
\label{fig:falsepeaks}
\end{figure}

To estimate the potential false peak rate in our high S/N sample due to
lensing shape noise \citep{kubo07} we performed two tests.  In our first test
we created a mean map of the 100 Monte Carlo reconstructions described in $\S
\ref{sec:peaks}$.  Each pixel in this map is the average value of this pixel
over the individual 100 Monte Carlo map realizations.  An average
signal-to-noise ratio map is created by dividing this map by the mean of the
$\kappa_{\rm{rms}}$ map ($\kappa_{\rm{rms}}/10$) from $\S \ref{sec:peaks}$.
Peaks are detected in this average signal-to-noise ratio map using exactly the same
procedure used to analyze the original $\kappa-\rm{S/N}$ map.  The resulting
peak histogram is shown in Figure \ref{fig:falsepeaks} (Left).   We find that
above our high signal-to-noise ratio cut ($\nu>3.5$) three peaks are recovered.  In
a second test we created individual signal-to-noise ratio maps from each of the 100
Monte Carlo realizations.  Here each individual Monte Carlo map is divided by
the $\kappa_{\rm{rms}}$ map from $\S \ref{sec:peaks}$.  Peaks are detected in
each of these individual signal-to-noise ratio maps using the same detection
procedure described in $\S \ref{sec:peaks}$.  The resulting average histogram
is shown in Figure \ref{fig:falsepeaks} (Right).  Here we detect $\sim 1$
false peaks above our high S/N cut. 
Both of the above tests indicate a potential false peak rate due to shape
noise of 1-3 (or $\sim 10-25\%$) above our high signal-to-noise ratio
cut $(\nu>3.5)$.  This percentage compares reasonably well with an estimate based on the proximity of our peaks to overdensities of galaxies  in physical and redshift space.  Based on the analysis of the environment near the peaks, we see 2-3 (16-25\%) of the structures appear not associated with or at $\gtrsim 3\arcmin$ from identified galaxy concentrations.  

In the above tests we do not simulate the signal due to uncorrelated large
scale structure along the line of sight which will increase the number of
detected objects that are falsely interpreted as cluster candidate. The
result of these tests therefore are only an estimate to the lower bound of the
false peak rate.  We emphasize that although false peaks are expected above
our high signal-to-noise ratio cut, the majority of our peaks appear to be
associated with real clusters which also appear to trace the large-scale
filamentary structure. 

\begin{figure}
\epsscale{1.0}
\plotone{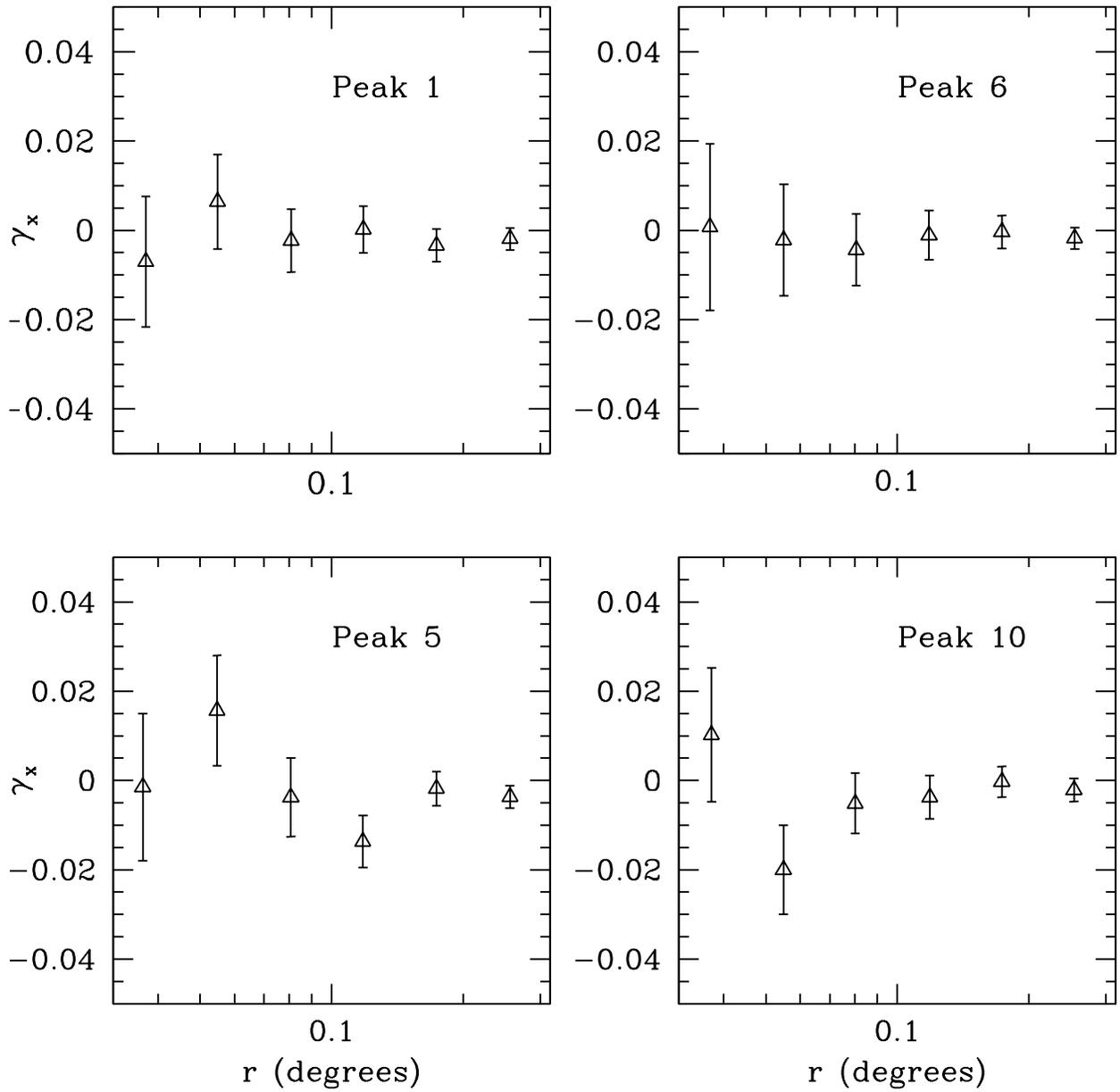}
\caption{Radial profile of the B-mode component centered on four high-signficance peaks in our convergence map.  The B-modes are all consistent with zero signal, indicating that the residuals in the PSF correction are not introducing a spurious signal near each lensing peak.}
\label{fig:bmode}
\end{figure}

We note here that residuals in the PSF correction could also potentially cause false peaks.  In \citet{kaiser93} type convergence maps this can be probed at some level by creating a B-mode map where source galaxies are rotated by 45 degrees relative to each grid point (e.g. \citet{massey07}).  One limitation of our maximum likelihood algorithm however is that in its current implementation it cannot create a B-mode map.  To probe PSF systematics we have alternatively measured radial shear B-modes for our high significance peaks (Figure \ref{fig:bmode}).  In the figure we show profiles for four of these peaks spread over the structures in the F2 field.  In each panel the profile is center on the peak coordinates given in Table \ref{tab:clusters}.  The measured B-mode for each peak is consistent with zero, indicating that there is no residual contamination coming our PSF correction.  Profiles for the other high significance peaks in the field are similarly consistent with a zero B-mode signal.

\subsection{Total S/N}
\label{sec:totsn}
\begin{figure}
\epsscale{1.0}
\plotone{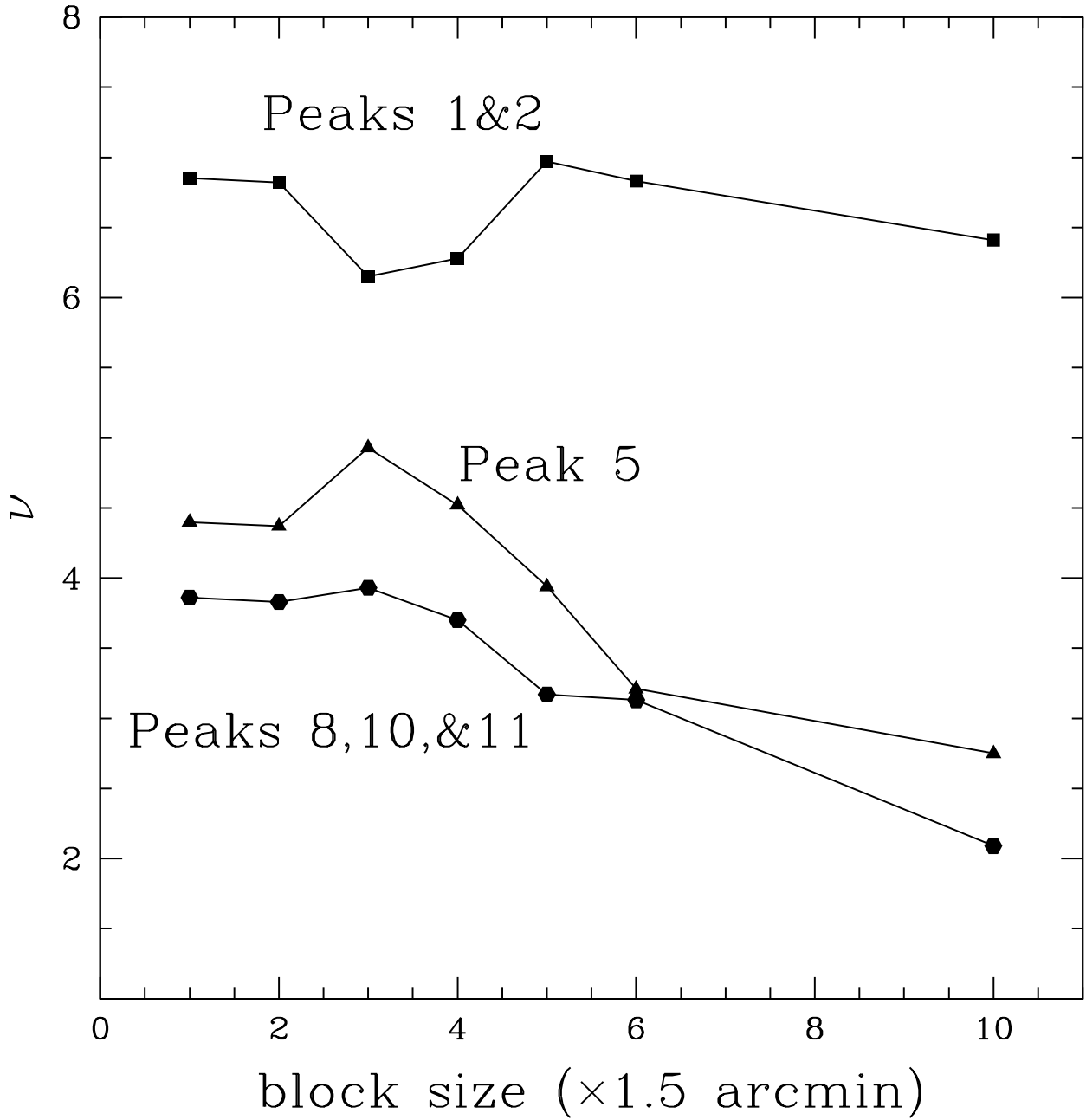}
\caption{Examples of total signal-to-noise ratio estimation using the method outlined in $\S \ref{sec:discussion}$.  The top curve is for the peak grouping 1 \& 2, the middle curve is for peak 5, and the bottom curve is for the peak grouping 8, 10, \& 11.  In each, the peak signal-to-noise ratio ($\nu$) is measured for different levels of block size after block averaging the $\kappa$ and individual Monte Carlo maps. $\kappa-\rm{S/N}$ maps are then created for each level of block size.  The curve for Peak 5, an isolated peak in our map, exhibits a clear maximum.  For the two other curves the maximum is not as well defined due to the proximity of nearby peaks, however an estimate for each peak grouping can still be obtained.  As expected the total signal-to-noise ratio values are higher than our estimates using only the peak signal-to-noise ratio from our $\kappa-\rm{S/N}$ map.}
\label{fig:block}
\end{figure}

To be consistent with previous studies we have selected peaks from the
$\kappa-\rm{S/N}$ map as described in $\S \ref{sec:peaks}$.  This provides a
good method to globally select peaks from the map, however because our
reconstruction technique does not track pixel-to-pixel correlations, neither
the peak nor the integrated signal-to-noise ratio are good estimates of total
signal-to-noise ratio of detection.  To calculate the total signal-to-noise ratio for
peaks we use a different approach where the original $\kappa$ map and the individual
Monte Carlo maps are block averaged at different levels of block size.
$\kappa-\rm{S/N}$ maps at each block size are then created in the same manner
used to create the original $\kappa-\rm{S/N}$ map $(\S \ref{sec:peaks})$.  For
small values of block size, due to the pixel-to-pixel correlations in the
$\kappa$ and $\kappa_{\rm{rms}}$ maps, both the signal and noise increase
linearly, hence the peak signal-to-noise ratio ($\nu$) will stay relatively
constant.  When the block size becomes larger than the noise correlation
length, noise will increase proportional to square root of the block size;
therefore, the peak signal-to-noise ratio will grow as more $\kappa$ pixels are used
in the average.  At some level of block size the peak signal-to-noise ratio will be
maximized when the pixel size matches the characteristic size of the cluster.  The peak
signal-to-noise ratio at this block size can then be used to estimate the total
signal-to-noise ratio of detection.  Past the maximum as the block size continues to
increase we expect that the signal-to-noise ratio will decrease as more noise pixels
are added.  An example of this technique is shown in Figure \ref{fig:block};
here we show peak S/N vs. block size curves for three different peaks in our
map.  The top curve is for the peak grouping 1 \& 2, the middle curve is for
peak 5, and the bottom curve is for the peak grouping of 8, 10, \& 11.  In
each of the curves the peak S/N reaches a maximum at certain level of block
size, and the value of the peak S/N at that block size is used to approximate
the total signal-to-noise ratio of detection.  This provides a near-optimal measure
of the integrated signal-to-noise ratio for isolated clusters, subject to the
constraint of finite pixel sizes.  One limitation of our total signal-to-noise ratio
estimation is that for large block sizes some peaks can become blended with
other peaks.  This behavior is exhibited in the top curve (Peaks 1 \& 2) and
the bottom curve (Peaks 8, 10, \& 11).  In such cases we can only estimate the
total signal-to-noise ratio for the combined system.  We list total S/N values using
this method for the remaining peaks in Table \ref{tab:clusters}.  

Our method of estimating the total signal-to-noise ratio for lensing peaks is qualitatively similar to previously-used techniques  in this field. For example, the P-statistics introduced by Schirmer et al. (2007) use variations in the scale of the kernel to effectively identify peaks based on their signal at a smoothing scale matching the peak size, even though they do not take the additional step of identifying the S/N at this kernel with the total S/N of the peak.  Our method is also similar to the method used in \citet{kaiser95}; however in that paper, the method was used to estimate the S/N ratio of faint galaxies rather than in estimating the significance of the lensing signal itself.

\subsection{The Abell 781 Complex}
\label{sec:a781}
As mentioned in $\S \ref{sec:clusters}$ there is a known X-ray cluster
associated with the western portion of Abell 781 complex that does not appear
in our reconstruction.  Our lensing map does hint at a peak at this position
but this peak appears to occupy only 1-2 pixels in our map.  We recently reported
a $1\sigma-2\sigma$ lensing detection of this cluster from an NFW fit in \citet{sehgal08}.  The
lensing significance of this cluster is low compared to what is expected from the X-ray mass
of this cluster \citep{sehgal08}.  The low lensing significance could
possibly be due to a remaining systematic, or could potentially point toward
an interesting physical effect which causes X-ray and lensing masses to be discrepant.  Observations of this region with a different telescope/imager are planned, and a detailed analysis will be presented in a future paper.

\subsection{A Candidate Void?}
\label{sec:void}
In $\S \ref{sec:peaks}$ we mentioned the detection of a candidate underdensity
in the $\kappa-{\rm{S/N}}$ map with a significant negative signal-to-noise ratio.  Strictly speaking, this region is a relative underdensity since it is underdense relative to other structure in the map.  Our ability to detect a relative undersity is unaffected by the mass sheet degeneracy transformation \citep{schneider95}.  The undersity is located at $\rm{R.A.}=09^{\rm{h}}20^{\rm{m}}22.7^{\rm{s}}$, $\rm{decl.}=+29^{\circ}54\arcmin03.5\arcsec$ and is detected with a peak significance $\nu\sim5$ and a total signal-to-noise ratio of $\sim5.5$.  From our lensing reconstruction alone we cannot directly confirm that this underdensity is in fact a void.  A galaxy redshift survey, an alternative mapping of structure, is a well known method of directly tracing underdense regions or voids \citep{geller89}.  We are conducting such a survey in the F2 field, the SHELS redshift survey \citep{geller05}, which should provide additional insights into this underdense region.  Confirmation of an underdensity detected in a lensing reconstruction would demonstrate that lensing maps can trace mass underdensities, in addition to overdensities.  We note that an excess of negative peaks was reported in \citet{miyazaki02}, however this excess has not been confirmed with a redshift survey.

\section{Summary}
\label{sec:summary}
In this work we have presented the application of a regularized maximum likelihood weak lensing
reconstruction to the Deep Lens Survey F2 field.  Our lensing
reconstruction reveals several projected structures in the F2 field which
appear to lie along a network of potential filaments.  In the southwest
section of our map we detect a structure which appears to lie at a redshift
$z\sim0.5$. Another structure, the Eastern-most structure in our
reconstruction, contains the confirmed cluster Abell 781 and appears to extend
southward for over a degree.  Known optical clusters at three redshifts
$z\sim0.3, 0.43, 0.5$ are potentially associated with significant lensing
peaks along this structure. 
Near the center of our map we detect another significant peak at a redshift
$z\sim0.32$.  The cluster associated with this lensing peak lies at a roughly
the same redshift as many of the peaks in the Eastern structure, however it is
unclear if this structure is physically associated with any part of the
Eastern structure.  These two structures also appear to surround a relative
underdense region with $\nu\sim5$, and total signal-to-noise ratio $\sim5.5$. 

The noise in our reconstruction is estimated from 100 individual Monte Carlo
realizations of the source galaxy catalog.  We find that generating Monte
Carlo $\kappa$ maps with this reconstruction technique is computationally
expensive, but does allow us to create a statistically correct
$\kappa-\rm{S/N}$ map.  Monte Carlo realizations also aid in estimating the
number of expected false peaks due to lensing shape noise.  We performed two
tests here using the Monte Carlo $\kappa$ maps; in one test we created an
average Monte Carlo significance map, and in another test we created
individual Monte Carlo significance maps and averaged the results.  Both of
these tests estimate $10-25\%$ false positive detections due to shape noise in our high
peak signal-to-noise ratio sample.  This is comparable to the number of peaks in our dataset which lack close counterparts in galaxy overdensities.

We have also presented a new method to estimate the total signal-to-noise ratio of
detection for lensing peaks.  Because pixel-to-pixel correlations are not
traced in our lensing reconstruction (or in any traditional direct
reconstruction) the peak signal-to-noise ratio underestimates the total lensing
significance.  Our method of estimating the total lensing significance uses
block averaging of the $\kappa$ map and the individual Monte Carlo maps to
create $\kappa-\rm{S/N}$ maps with different levels of block size.  The peak
signal-to-noise ratio will become maximized here when the peak matches the characteristic size of
the cluster, and this peak signal-to-noise ratio can be used to estimate the total
lensing significance.  We find that this method works quite well for isolated
peaks, but for a grouping of peaks only the significance of the combined group can
be measured. 

In future work, photometric redshifts of source galaxies in the F2 field could
allow us to make tomographic maps using this method and would also allow for
the determination of the mass of significant lensing peaks.  Magnification
information in the F2 field could also be incorporated into our
reconstruction, potentially breaking the mass sheet degeneracy
\citep{broadhurst95}.  We are also 
pursing optical cluster finding in the DLS, however this is beyond the scope
of our current study.

\acknowledgments

We would like to thank NOAO for generous allocations of telescope time for the
survey.  This work was supported by NSF grants AST-0134753 and AST-0708433.
Kitt Peak National Observatory, National Optical Astronomy Observatory, is
operated by the Association of Universities for Research in Astronomy (AURA)
under cooperative agreement with the National Science Foundation.  Fermilab is
operated by Fermi Research Alliance, LLC under Contract No. DE-AC02-07CH11359
with the United States Department of Energy.

\clearpage



\begin{deluxetable}{cccccc}
\tablecolumns{6}
\tablewidth{0pc}
\tablecaption{Lensing peaks with $\nu>3.5$ in the DLS F2 \label{tab:clusters} field}
\tablehead{
\colhead{Peak Rank} & \colhead{R.A. ($^{\circ}$)}  & \colhead{Decl. ($^{\circ}$)} & \colhead{$\nu$\tablenotemark{a}} & \colhead{$z$} & \colhead{$S/N_{t}$\tablenotemark{b}}}
\startdata
1 & 140.2294 & +30.4677 & 6.6 & 0.427, 0.291 & 7.0\\
2 & 140.0829 & +30.5070 & 5.7 & 0.302 & 7.0\\
3 & 140.3072 & +30.2299 & 5.3 & $\sim 0.5$ & 5.4\\
4 & 140.4161 & +29.6915 & 4.5 & N/A\tablenotemark{c} & 4.0\\
5 & 139.6471 & +29.8726 & 4.4 & 0.317 & 4.9\\
6 & 140.2994 & +29.5624 & 4.2 & N/A\tablenotemark{c} & 4.0\\
7 & 140.5350 & +30.1093 & 4.1 & --- & 3.6\\
8 & 138.9965 & +29.5569 & 3.9 & 0.53 & 3.9\\
9 & 140.4174 & +29.4782 & 3.9 & $\sim0.34$ & 4.0\\
10 & 139.0007 & +29.8431 & 3.8 & 0.55 & 3.9\\
11 & 138.9588 & +29.6458 & 3.7 & 0.53 & 3.9\\
12 & 140.5499 & +29.9367 & 3.6 & --- & 3.6\\
\tableline
void & 140.0946 & +29.9010 & $\sim5$ & --- & $\sim5.5$\\
\enddata
\tablenotetext{a}{$\nu$ is the peak signal-to-noise ratio measured from the $\kappa-\rm{S/N}$ map}
\tablenotetext{b}{$S/N_{t}$ is the total signal-to-noise ratio measured using the method outlined in $\S \ref{sec:totsn}$}
\tablenotetext{c}{Peaks 4 and 6 do not appear to be associated with galaxy overdensities and are likely false positives}
\tablecomments{Due to the resolution of our map, some peaks are blended together.  In this case we report the total signal-to-noise ratio of detection for the combined systems.}
\end{deluxetable}

\end{document}